\documentclass[12pt]{iopart}
\usepackage{graphicx}
\usepackage{citesort}
\usepackage{color}

\begin{document}

\title{Possible spin frustration in Nd$_2$Ti$_2$O$_7$ probed by muon spin relaxation}

\author{Hanjie Guo$^1$, Hui Xing$^1$, Jun Tong$^1$, Qian Tao$^1$, Isao Watanabe$^2$, and Zhu-an Xu$^1$}

\address{$^1$Department of Physics and State Key Laboratory of Silicon Materials, Zhejiang University, Hangzhou 310027, China}
\address{$^2$Advanced Meson Science Laboratory, RIKEN, 2-1 Hirosawa, Wako, Saitama 351-0198, Japan}

\ead{zhuan@zju.edu.cn}

\vspace{10pt}
\begin{indented}
\item[]September 2014
\end{indented}

\begin{abstract}
Muon spin relaxation on Nd$_2$Ti$_2$O$_7$ (NTO) and NdLaTi$_2$O$_7$ (NLTO) compounds are presented. The time spectra for both compounds are as expected for the paramagnetic state at high temperatures, but deviate from the exponential function below about 100 K. The muon spin relaxation rate increases with decreasing temperature firstly, and then levels off below about 10 K, which is reminiscent of the frustrated systems. An enhancement of the relaxation rate by a longitudinal field in the paramagnetic state is observed for NTO, and eliminated by a magnetic dilution for the NLTO sample. This suggests that the spectral density is modified by a magnetic dilution and thus indicates that the spins behave cooperatively rather than individually. Zero-field measurement at 0.3 K indicates that the magnetic ground state for NTO is ferromagnetic.
\end{abstract}

%
\section{Introduction}

Geometrically frustrated magnetic systems have been widely investigated in order to understand novel phenomena such as the spin ice \cite{Ramirez_spin_ice,Bramwell_Ho2Ti2O7}, spin liquid \cite{Gardner_Tb2Ti2O7} etc., due to the subtle competitions in the frustrated compounds.
A survey of the monopole state in spin ice by muon spin relaxation ($\mu$SR) has been controversial \cite{Castelnovo_monopole,Bramwell_monopole,Dunsiger_Dy2Ti2O7,Blundell_monopole}.
Typical geometrical frustrated structures consist of the kagome structure \cite{Uemura_kagome} and the pyrochlore structure with a formula $A_2B_2$O$_7$ \cite{Gardner_pyrochlore_oxide}, where $A$ and $B$ ions form two separative networks of corner-sharing tetrahedra and penetrate each other.
Frustration usually prevents a formation of a long range magnetically ordered state even with a large spin-spin interaction, and the ground state is highly degenerated.
Such a state can be termed as a ``cooperative paramagnetic" state as observed in Tb$_2$Ti$_2$O$_7$ (TTO) \cite{Gardner_Tb2Ti2O7}. $\mu$SR measurements on TTO reveal that Tb$^{3+}$ spins fluctuate down to 70 mK and the muon spin relaxation rate is temperature independent at low temperatures \cite{Gardner_Tb2Ti2O7}.

When $A$ is occupied by rare earth elements with large ionic radius, such as Nd$_2$Ti$_2$O$_7$ (NTO), the compound crystallizes in the monoclinic structure with space group $P112_1$ instead of the frustrated pyrochlore structure \cite{Shcherbakova_NTO,Scheunemann_structure,Ishizawa_structure}. The corner-sharing TiO$_6$ octahedra and Nd ions form slabs along the $b$ direction, making it easily cleaved along the $b$ direction due to the layered structure. The Nd ions are displaced along the $b^\ast$ direction from the geometrical centers of coordination polyhedra, forming zig-zag chains in the $ab$ plane.
A large Weiss temperature ($\theta_W$) of -42 K for a polycrystalline sample has been extracted while NTO shows paramagnetic behavior down to at least 2 K \cite{Xing_magnetic_susceptibility}.
Specific heat measurement exhibits a sharp peak at $T_0$ = 0.59 K \cite{Hui_ac}, indicating a possible magnetic ordering.  The frustration index defined as $f = |\theta_W|/T_0$ is thus as large as 70 \cite{Ramirez_index}, although frustration is not obvious in NTO from the crystal structure. This implies that an unusual spin dynamics may exist in the intermediate temperature range in NTO. Detailed studies in this system may shed new light on our understanding of the role of spin-spin interactions and single-ion effect in the frustrated systems.

In this paper, we show that muon spin relaxation measurements on NTO resemble that of many frustrated compounds.
The muon spin relaxation rate $\lambda$ increases with decreasing temperature, reflecting the slowing down of the fluctuating Nd$^{3+}$ moments.
However, $\lambda$ levels off and exhibits a plateau below about 10 K, which has been observed in many frustrated systems while a comprehensive understanding is still lacking. Strikingly, $\lambda$ is found to be enhanced by a longitudinal field in the paramagnetic state, which is not expected and gives a direct evidence of the enhancement of the spectral density at the muon Larmor frequency.
Such an enhancement is eliminated by magnetic dilution at the Nd site, suggesting that the spin-spin correlations between Nd$^{3+}$ moments should play a key role.

\section{Experiment}

Single crystals of NTO and NdLaTi$_2$O$_7$ (NLTO) were grown by the floating zone image furnace as reported by Xing \textit{et al}\cite{Xing_magnetic_susceptibility}.
The samples were cleaved naturally along the \textit{b} direction. $\mu$SR experiments were performed at the RIKEN-RAL muon facility at the Rutherford-Appleton Laboratory, U.K.
Plate-like samples were mounted on the high-purity silver sample holder. Spin-polarized muons were injected into the sample and the decayed positrons which were ejected preferentially along the muon spin direction were accumulated.
The initial muon spin polarization is in parallel with the beam line. Forward and backward counters are located in the upstream and downstream of the beam line, respectively. The time dependent asymmetry ($\mu$SR spectrum) of the muon spin polarization is defined as $A(t) = [F(t)-\alpha B(t)]/[F(t)+\alpha B(t)]$, where $F(t)$ and $B(t)$ are the muon events counted by the forward and backward counters, respectively.
Parameter $\alpha$ reflects the relative counting efficiency of the forward and backward counters. The experiments were performed in the zero-field (ZF) and longitudinal-field (LF) configurations, respectively. The LF was applied along the initial muon-spin polarization.

\section{Experimental results}

\begin{figure}
\centering
\includegraphics[width=14cm]{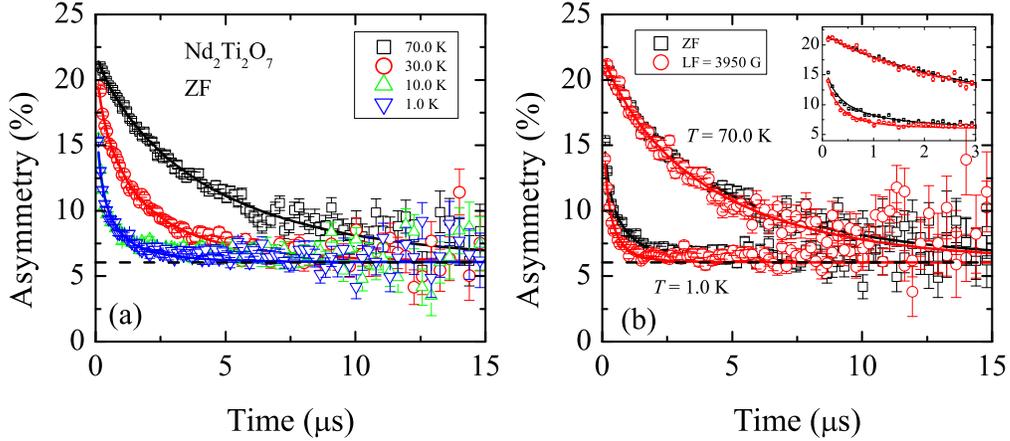}
\caption{(Color online) (a) Time spectra for Nd$_2$Ti$_2$O$_7$ measured at various temperatures in ZF. (b) Comparison of the time spectra measured in ZF and LF of 3950 G. Black square (red circle) is the spectrum in ZF (LF). Solid curves are fitted results according to equation (\ref{stretched_exponential}). The dashed line indicates the background signal from the silver sample holder. The inset of (b) shows the early time region where the relaxation rate in LF is larger that that in ZF when $T$ = 1.0 K.\label{Fig_NTO_ZF_LF_spectrum}}
\end{figure}

Typical time spectra for NTO measured in ZF at various temperatures are shown in figure \ref{Fig_NTO_ZF_LF_spectrum}(a).
The muon spin relaxation becomes faster with decreasing temperature. A loss of the initial asymmetry at $t$ = 0 is observed due to the large relaxation rate at low temperatures. The spectra overlap with each other below about 10 K,
indicating similar relaxation rates.
A comparison of the spectra measured in ZF and LF of 3950 G at 70.0 and 1.0 K, respectively, are displayed in figure \ref{Fig_NTO_ZF_LF_spectrum}(b). No significant difference is observed at 70.0 K,
as expected for a paramagnetic state. Strikingly, the spectrum at 1.0 K decays faster in the LF than that in the ZF condition, as seen from the inset of figure \ref{Fig_NTO_ZF_LF_spectrum}(b),
which is in contrast to a general speculation that the muon spin polarization will be recovered by an LF,
or at least unchanged if the fluctuation rate of the internal fields is much larger than the muon Larmor frequency in the paramagnetic state.

In order to obtain more insight on the spin dynamics, we tried different functions to fit the spectra in figure \ref{Fig_NTO_ZF_LF_spectrum}.
It is found that a simple exponential function is not applicable to simulate all the spectra, especially at low temperatures. On the other hand,
all the spectra can be fitted using the stretched-exponential function
\begin{equation}\label{stretched_exponential}
    A(t) = A_S\mathrm{exp}[-(\lambda t)^\beta]+A_{B},
\end{equation}
where $A_S$ and $A_B$ are the amplitudes of the signal from the sample and sample holder, respectively. $A_B$ can be regarded as time independent and was determined by the low temperature spectrum exhibiting fast relaxation. The dashed line in figure \ref{Fig_NTO_ZF_LF_spectrum} indicates the value of $A_B$.
$A(0)$ is the initial asymmetry and fixed to the value measured at high temperatures.
Parameters $\lambda$ and $\beta$ are the muon spin relaxation rate and the stretched exponent, respectively.
When $\beta$ is smaller than 1, it suggests a distribution of the relaxation rate. The fitted value of $\lambda$ is only a typical relaxation rate, and the majority of the spectral weight in the relaxation rate will move to much lower values when $\beta$ is small \cite{Johnston_stretched}.

\begin{figure}
\centering
\includegraphics[width=10cm]{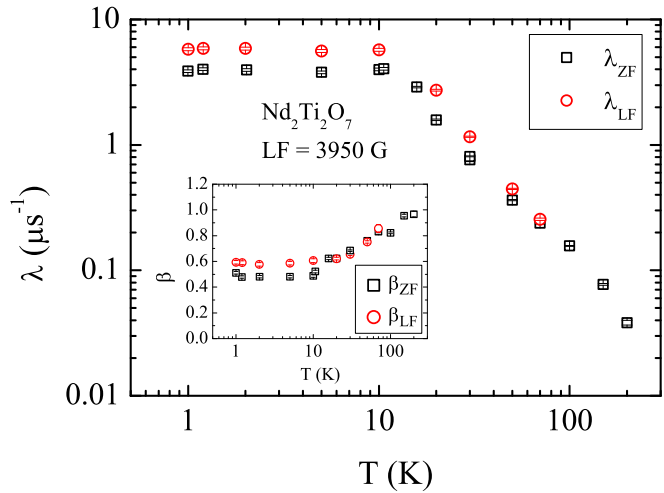}
\caption{(Color online) Temperature dependence of the extracted relaxation rate $\lambda$ for Nd$_2$Ti$_2$O$_7$. The inset shows the result of exponent parameter $\beta$. Black square (red circle) is the result obtained in ZF (LF) condition.\label{NTO_parameter}}
\end{figure}

The extracted temperature dependence of $\lambda$ and $\beta$ are displayed in figure \ref{NTO_parameter}.
The subscript ZF (LF) indicates that the measurements were performed in the ZF (LF) conditions.
Both $\lambda_\mathrm{ZF}$ and $\lambda_\mathrm{LF}$ increases firstly with a decrease in temperature. Below about 10 K, however, both become temperature independent and $\lambda_\mathrm{LF}$ is about 2 $\mu$s$^{-1}$ larger than $\lambda_\mathrm{ZF}$.
From the inset of figure \ref{NTO_parameter}, it can be seen that both $\beta_\mathrm{ZF}$ and $\beta_\mathrm{LF}$ are approaching to 1 at high temperatures, indicating a homogeneous paramagnetic state.
With the decrease in temperature, $\beta$ decreases gradually and shows temperature independent behavior below about 10 K with a value of about 0.5 and 0.6 for $\beta_\mathrm{ZF}$ and $\beta_\mathrm{LF}$, respectively. It is worth noting that the $\beta$ value is close to the expected value, 0.5, in the extremely diluted magnetic system \cite{Uemura_spin_glass}.

\begin{figure}
  \centering
  \includegraphics[width=10cm]{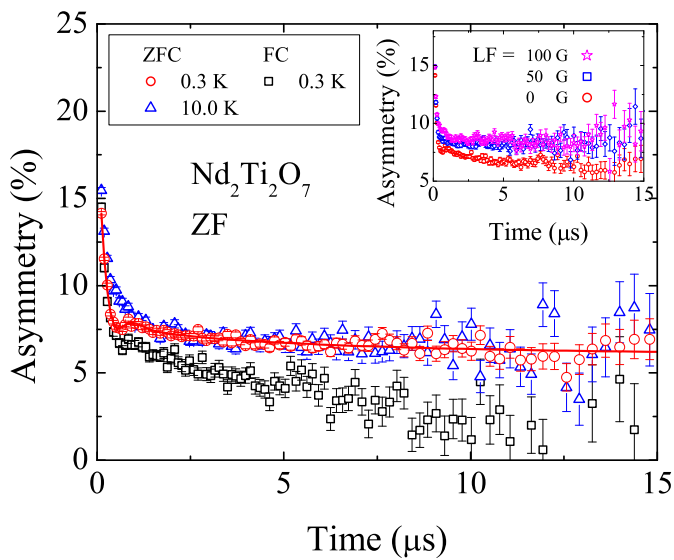}\\
  \caption{(Color online) Time spectra for Nd$_2$Ti$_2$O$_7$ measured down to 0.3 K in ZF. The black square (red circle) spectrum was measured after a FC (ZFC) process. The magnitude of LF during the FC process is 3950 G. The inset shows the longitudinal field dependence of the spectrum at 0.3 K.}\label{Fig_NTO_base_spectrum}
\end{figure}

Figure \ref{Fig_NTO_base_spectrum} shows the spectra measured in ZF after zero-field-cooling (ZFC) and field-cooling (FC) processes down to 0.3 K. The time spectrum at 10 K is shown as well for comparison.
It can be observed that the spectra exhibit a bifurcation after around 0.5 $\mu$s between the FC and ZFC measurements.
The spectrum after the ZFC process is observed to exhibit a dip around 0.5 $\mu$s and then the asymmetry is recovered. The LF measurements after ZFC process shown in the inset of figure \ref{Fig_NTO_base_spectrum} exhibit a decoupling behavior.
These behaviors indicate an appearance of static internal fields at the muon site,
consistent with the specific heat measurement that implies a magnetic ordering below about 0.59 K.
Therefore, the discrepancy between the ZFC and FC measurements at 0.3 K may be due to the residual magnetization and the stray field to the silver sample holder, which then induces the relaxation of the background signal.
The hysteresis effect is not found above $T_0$ as expected for a paramagnetic state. Such a hysteresis effect suggests that NTO is ferromagnetic at the ground state. The mean-field exchange interactions along three axes were estimated to be positive, which may induce the ferromagnetic ordering \cite{Xing_magnetic_susceptibility}.

The spectrum at 0.3 K is similar to the case of highly disordered magnets which are simulated by a static ``Gaussian-broadened Gaussian" Kubo-Toyabe function \cite{Noakes_GBG}. It also resembles that of exponential distribution of fields in the spin-ladder system Sr(Cu$_{1-x}$Zn$_x$)$_2$O$_3$ \cite{Larkin_ZF}.
Here, we notice that the fast initial relaxation within 0.5 $\mu$s can be traced to the paramagnetic state.
From the inset of figure \ref{Fig_NTO_base_spectrum}, the late-time spectrum is decoupled by an LF as expected for the static internal fields and no significant difference is observed between the spectra measured at 50 and 100 G, while the early-time region is almost unchanged. Considering that 0.3 K is very close to $T_0$, these observations suggest a coexistence of static and dynamic fields at 0.3 K.
Therefore, in order to obtain the value of internal fields, we use a simpler phenomenological function as follows:
\begin{eqnarray}\label{Eq_order}
  A(t) = A_1\mathrm{e}^{-\lambda_1t} + A_2\mathrm{cos}(\gamma_\mu B_\mu t+\varphi)\mathrm{e}^{-\lambda_2 t}+A_3\mathrm{e}^{-(\lambda t)^\beta}+A_B,
\end{eqnarray}
where $\lambda_1$ represents the dynamic relaxation rate, $B_\mu$ the mean value of internal fields at the muon site, $\varphi$ and $\lambda_2$ are the phase and damping rate of the muon spin precession, respectively, and $\gamma_\mu/2\pi$ = 13.55 kHz/G is the gyromagnetic ratio of muon. The third term accounts for the paramagnetic component. The fitted result is the red curve shown in figure \ref{Fig_NTO_base_spectrum}. The obtained $B_\mu$ is 63.3 $\pm$ 5.2 G, $\lambda_1$ is 0.152 $\pm$ 0.052 $\mu$s$^{-1}$, $\lambda_2$ is 3.52 $\pm$ 0.60 $\mu$s$^{-1}$, and $\lambda$ and $\beta$ are 6.9 $\pm$ 1.5 $\mu$s$^{-1}$ and 0.52 $\pm$ 0.15, respectively.

\begin{figure}
\centering
\includegraphics[width=14cm]{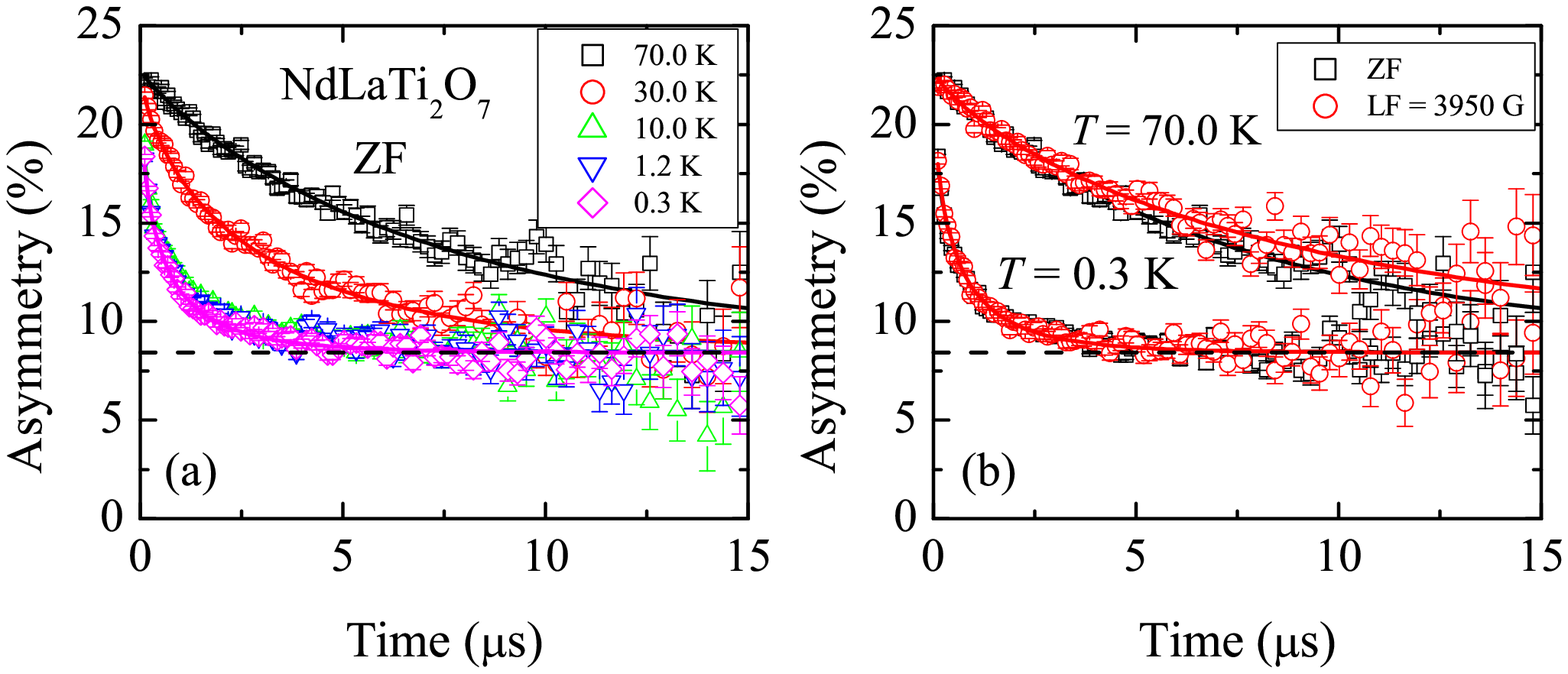}
\caption{(Color online) (a) Time spectra for NdLaTi$_2$O$_7$ measured at various temperatures in ZF. (b) Comparison of the time spectra measured in ZF and LF of 3950 G. Black square (red circle) is the spectrum in ZF (LF). Solid curves are fitted results according to equation (\ref{stretched_exponential}). The dashed line indicates the background signal from the silver sample holder. \label{NLTO_ZF_LF_spectrum}}
\end{figure}

\begin{figure}
  \centering
  \includegraphics[width=10cm]{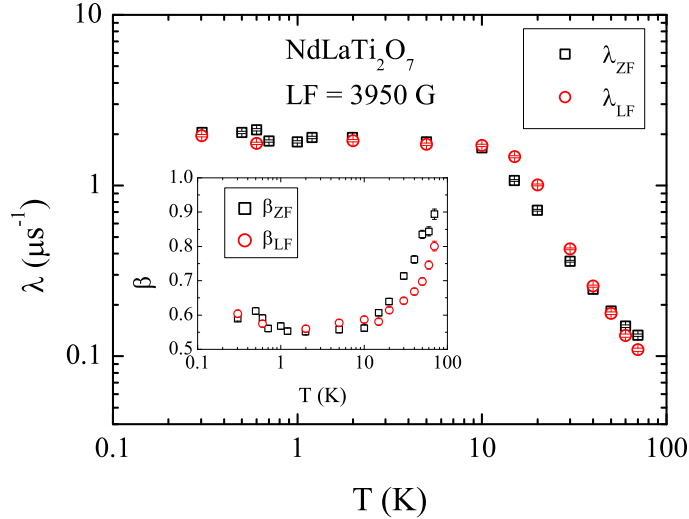}\\
  \caption{(Color online) Temperature dependence of the extracted relaxation rate $\lambda$ for NdLaTi$_2$O$_7$. The inset shows the result of exponent parameter $\beta$. Black square (red circle) is the result obtained in ZF (LF).}\label{NLTO_parameter}
\end{figure}

We also performed $\mu$SR experiments on the La-doped NLTO, as shown in figure \ref{NLTO_ZF_LF_spectrum}.
The ZF spectra show similar depolarization behavior as NTO, and no static signal is detected down to 0.3 K.
The spectra recorded in an LF of 3950 G are shown in figure \ref{NLTO_ZF_LF_spectrum}(b). In contrast to the case of NTO, the muon spin relaxation behavior is nearly unchanged by the LF both at high and low temperatures.

The extracted temperature dependence of $\lambda$ and $\beta$ for NLTO are shown in figure \ref{NLTO_parameter}.
Compared with the NTO case, both $\lambda_\mathrm{ZF}$ and $\lambda_\mathrm{LF}$ in NLTO match each other quite well in the measured temperature range and saturate below about 10 K. The temperature from which $\lambda$ levels off is indistinguishable from that of NTO in our current study due to the lack of data points. The saturated $\lambda_\mathrm{ZF}$ is about 2 $\mu$s$^{-1}$ smaller than that of NTO below about 10 K, reflecting the dilution effect of the Nd$^{3+}$ moments.

\section{Discussions}

We first clarify the validity of equation (\ref{stretched_exponential}) used in the paramagnetic state. In the motional narrowing limit, i.e., when $\gamma_\mu \sqrt{\langle B_\mu^2\rangle}\tau_c \ll 1$, where $\tau_c$ is the correlation time of the internal field,
the muon spin relaxation can be expressed as an exponential function \cite{Hayano}.
The relaxation rate can be expressed as
\begin{equation}\label{Eq_Redfield}
  \lambda = \frac{2\gamma_\mu^2\Delta^2 \tau_c}{1 + \omega_L^2\tau_c^2},
\end{equation}
where $\omega_L = \gamma_\mu B_\mathrm{LF}$ is the muon Larmor frequency, and $\langle B_x^2\rangle = \langle B_y^2\rangle = \Delta^2$ is assumed.
A root-exponential function is derived assuming different muon sites at which muons sense internal fields with different distribution width, as in the spin glass case \cite{Uemura_spin_glass}. The stretched exponential function is often used as a generalized case. If we consider equation (\ref{Eq_Redfield}) for simplicity,
and take $B_\mu$ = 63.3 G at 0.3 K as $\Delta$, it can be extracted that $\tau_c \sim$ 7$\times 10^{-8}$ s below 10 K in the ZF case. Therefore, $(\gamma_\mu \Delta \tau_c)^{-1} \sim$ 3, which satisfies the narrowing condition \cite{Reotier_narrowing}.

We next discuss the magnetic field effect on the spin dynamics in NTO.
In ZF or when $\omega_L \tau_c \ll 1$, $\lambda \sim 2\gamma_\mu^2 \Delta^2 \tau_c$, the increase of $\lambda$ with decreasing temperature above $\sim$10 K reflects the slowing down of the fluctuating Nd$^{3+}$ moments. According to equation (\ref{Eq_Redfield}), the magnetic field will suppress the relaxation rate if the magnetic field does not affect $\tau_c$.
This expectation disagrees with the experimental results below about 10 K.
The magnetic field is observed to enhance the muon spin relaxation rate from $\sim$4 to 6 $\mu$s$^{-1}$ by an LF of 3950 G. Such a behavior is quite rare in the paramagnetic state but not unprecedented. In Tb$_2$Sn$_2$O$_7$,
the enhancement of the muon spin relaxation rate by an LF is observed both in the magnetically ordered state and paramagnetic state, and is accounted for by the magnetic field enhanced density of magnetic excitation at low energies \cite{Rotier_magnetic_field}. As shown in figure \ref{Fig_NTO_base_spectrum}, the ZF-$\mu$SR measurement at 0.3 K suggests a ferromagnetic ground state in NTO. Short range ferromagnetic fluctuations can exist at higher temperatures and the ferromagnetic correlation time can be enhanced by the applied LF; thus the muon spin relaxation rate is enhanced.

As discussed in previous literatures \cite{Slichter_book,Dunsiger_Gd2Ti2O7}, the muon spin relaxation rate $\lambda = 1/T_1$ is proportional to the spectral density $J(\omega)$ at the muon Larmor frequency $\omega_L$. $J(\omega)$ is proportional to the Fourier transformation of the correlation function of internal field $\langle B_q(t)B_q(t+\tau)\rangle$.
Therefore, the increase of $\lambda$ by applying a magnetic field below $\sim$10 K gives a direct evidence that the magnetic field enhances the spectral density $J(\omega)$ in the paramagnetic state, which is not assumed in previous studies in the paramagnetic state of spin glass \cite{Keren_1996,Keren_2001}.
In the frustrated SrCr$_{9p}$Ga$_{12-9p}$O$_{19}$, the magnetic field dependent $\lambda$(H) was observed to follow the same behavior regardless when $p$ is above or below the percolation threshold,
indicating that the spectral density is unchanged by magnetic dilution; therefore, the spins fluctuate individually \cite{Keren_SCGO}.
In the NTO system, $\lambda$ is enhanced by an LF in NTO while such an effect is not observed in NLTO, suggesting that the spectral density is modified by the magnetic dilution.
Thus, we speculate that the Nd$^{3+}$ moments behave cooperatively instead of individually.

We also note that $\lambda$ becomes temperature independent below $\sim$10 K both in ZF and LF conditions. Such a behavior has been observed in frustrated systems with a variety of ground states, e.g., Tb$_2$Ti$_2$O$_7$ in the spin liquid state \cite{Gardner_Tb2Ti2O7}, Dy$_2$Ti$_2$O$_7$ in the spin ice state \cite{Dunsiger_Dy2Ti2O7}, Gd$_2$Ti$_2$O$_7$, and Tb$_2$Sn$_2$O$_7$ in a magnetically ordered state \cite{Dunsiger_Gd2Ti2O7,Bert_Tb2Sn2O7}, and the kagome-like volborthite \cite{Fukaya_volborthite,Bert_volborthite}.
Although geometrical frustration is not obvious in NTO,
it has to be mentioned that the nearest-neighbor (NN) and next-nearest-neighbor (NNN) interactions may compete with each other, as the $J_1$-$J_2$ model in a two dimensional square lattice \cite{Lacroix_book}.
At present, the dominant NN exchange interaction can be estimated to be positive \cite{Xing_magnetic_susceptibility}, while the sign of the NNN exchange interaction is still unknown, further studies are necessary.

The $\mu$SR spectral shape for NTO in the paramagnetic state is unusual.
As shown in the inset of figure \ref{NTO_parameter}, the parameter $\beta$ begins to deviate from 1 below $\sim$100 K, and decreases with decreasing temperature. The $\beta$ value levels off below $\sim$10 K with $\beta_\mathrm{ZF}$ $\sim$ 0.5.
In a spin glass system with extremely diluted inhomogeneous magnetic moments, the $\mu$SR spectrum is expected to be root exponential, i.e., $\beta$ = 0.5 \cite{Uemura_spin_glass}.
One possibility of the inhomogeneity in NTO is the oxygen deficiency or different oxygen sites because muon tends to locate near the oxygen and form a $\mu^+$-O$^{2-}$ bond.
The high temperature homogeneity, i.e., $\beta$ close to 1, may be due to the muon diffusion in the sample and smears out the local variance. On the other hand, we notice that the distances between oxygens and their nearest Nd$^{3+}$ ions are quite close \cite{Ishizawa_structure}.
Considering the dense magnetic moments compared to the spin glass case, the internal field distributions should be quite different.
In such a case, the spectral shape of the current study may have different origins and deserves further investigations.

\section{Summary}

ZF-$\mu$SR measurements at the base temperature suggest that the ground state of NTO is ferromagnetic. The temperature dependent behavior of $\lambda$ for NTO in the paramagnetic state resembles that of many frustrated materials, implying that possible competition between the NN and NNN exchange interaction may exist and needs more studies. The modification of the spectral density by a magnetic dilution indicates that the Nd$^{3+}$ spins behave cooperatively rather than individually.

\ack{We thank F. L. Ning for useful discussions. This work is supported by the National Basic Research Program of China (Grant No. 2012CB821404), the National Science Foundation of China (Grant No. U1332209), and the Fundamental Research Funds for the Central Universities of China.}

\section*{References}
\bibliography{Nd2Ti2O7}

\providecommand{\newblock}{}
\begin{thebibliography}{10}
\expandafter\ifx\csname url\endcsname\relax
  \def\url#1{{\tt #1}}\fi
\expandafter\ifx\csname urlprefix\endcsname\relax\def\urlprefix{URL }\fi
\providecommand{\eprint}[2][]{\url{#2}}

\bibitem{Ramirez_spin_ice}
Ramirez A~P, Hayashi A, Cava R~J, Siddharthan R and Shastry B~S 1999 {\em
  Nature\/} {\bf 399} 333--335

\bibitem{Bramwell_Ho2Ti2O7}
Bramwell S~T, Harris M~J, den Hertog B~C, Gingras M~J~P, Gardner J~S, McMorrow
  D~F, Wildes A~R, Cornelius A~L, Champion J~D~M, Melko R~G and Fennell T 2001
  {\em Phys. Rev. Lett.\/} {\bf 87} 047205

\bibitem{Gardner_Tb2Ti2O7}
Gardner J~S, Dunsiger S~R, Gaulin B~D, Gingras M~J~P, Greedan J~E, Kiefl R~F,
  Lumsden M~D, MacFarlane W~A, Raju N~P, Sonier J~E, Swainson I and Tun Z 1999
  {\em Phys. Rev. Lett.\/} {\bf 82} 1012

\bibitem{Castelnovo_monopole}
Castelnovo C, Moessner R and Sondhi S~L 2008 {\em Nature\/} {\bf 451} 42--45

\bibitem{Bramwell_monopole}
Bramwell S~T, Giblin S~R, Calder S, Aldus R, Prabhakaran D and Fennell T 2009
  {\em Nature\/} {\bf 461} 956--959

\bibitem{Dunsiger_Dy2Ti2O7}
Dunsiger S~R, Aczel A~A, Arguello C, Dabkowska H, Dabkowski A, Du M~H, Goko T,
  Javanparast B, Lin T, Ning F~L, Noad H~M~L, Singh D~J, Williams T~J, Uemura
  Y~J, Gingras M~J~P and Luke G~M 2011 {\em Phys. Rev. Lett.\/} {\bf 107}
  207207

\bibitem{Blundell_monopole}
Blundell S~J 2012 {\em Phys. Rev. Lett.\/} {\bf 108} 147601

\bibitem{Uemura_kagome}
Uemura Y~J, Keren A, Kojima K, Le L~P, Luke G~M, Wu W~D, Ajiro Y, Asano T,
  Kuriyama Y, Mekata M, Kikuchi H and Kakurai K 1994 {\em Phys. Rev. Lett.\/}
  {\bf 73} 3306

\bibitem{Gardner_pyrochlore_oxide}
Gardner J~S, Gingras M~J~P and Greedan J~E 2010 {\em Rev. Mod. Phys.\/} {\bf
  82} 53

\bibitem{Shcherbakova_NTO}
Shcherbakova L~G, Mamsurova L~G and Sukhanova G~E 1979 {\em Russ. Chem. Rev.\/}
  {\bf 48} 228

\bibitem{Scheunemann_structure}
Scheunemann K and M¨¹ller-Buschbaum H 1975 {\em J. Inorg. Nucl. Chem.\/} {\bf
  37} 2261--2263

\bibitem{Ishizawa_structure}
Ishizawa N, Ninomiya K, Sakakura T and Wang J 2013 {\em Acta Cryst. E\/} {\bf
  69} i19

\bibitem{Xing_magnetic_susceptibility}
Xing H, Long G, Guo H, Zou Y, Feng C, Cao G, Zeng H and Xu Z~A 2011 {\em J.
  Phys.: Condens. Matter\/} {\bf 23} 216005

\bibitem{Hui_ac}
Xing H {\em data unpublished\/}

\bibitem{Ramirez_index}
Ramirez A~P 1994 {\em Annu. Rev. Mater. Sci.\/} {\bf 24} 453--480

\bibitem{Johnston_stretched}
Johnston D~C 2006 {\em Phys. Rev. B\/} {\bf 74} 184430

\bibitem{Uemura_spin_glass}
Uemura Y~J, Yamazaki T, Harshman D~R, Senba M and Ansaldo E~J 1985 {\em Phys.
  Rev. B\/} {\bf 31} 546

\bibitem{Noakes_GBG}
Noakes D~R and Kalvius G~M 1997 {\em Phys. Rev. B\/} {\bf 56} 2352--2355

\bibitem{Larkin_ZF}
Larkin M~I, Fudamoto Y, Gat I~M, Kinkhabwala A, Kojima K~M, Luke G~M, Merrin J,
  Nachumi B, Uemura Y~J, Azuma M, Saito T and Takano M 2000 {\em Physica B\/}
  {\bf 289-290} 153--156

\bibitem{Hayano}
Hayano R~S, Uemura Y~J, Imazato J, Nishida N, Yamazaki T and Kubo R 1979 {\em
  Phy. Rev. B\/} {\bf 20} 850

\bibitem{Reotier_narrowing}
Dalmas~de R\'{e}otier P and Yaouanc A 1997 {\em J. Phys.: Condens. Matter\/}
  {\bf 9} 9113

\bibitem{Rotier_magnetic_field}
Dalmas~de R\'{e}otier P, Yaouanc A, Keller L, Cervellino A, Roessli B, Baines
  C, Forget A, Vaju C, Gubbens P~C~M, Amato A and King P~J~C 2006 {\em Phys.
  Rev. Lett.\/} {\bf 96} 127202

\bibitem{Slichter_book}
Slichter C~P 1990 {\em Principles of Magnetic Resonance\/} 3rd ed (Springer)

\bibitem{Dunsiger_Gd2Ti2O7}
Dunsiger S~R, Kiefl R~F, Chakhalian J~A, Greedan J~E, MacFarlane W~A, Miller
  R~I, Morris G~D, Price A~N, Raju N~P and Sonier J~E 2006 {\em Phys. Rev. B\/}
  {\bf 73} 172418

\bibitem{Keren_1996}
Keren A, Mendels P, Campbell I~A and Lord J 1996 {\em Phys. Rev. Lett.\/} {\bf
  77} 1386--1389

\bibitem{Keren_2001}
Keren A, Bazalitsky G, Campbell I and Lord J~S 2001 {\em Phys. Rev. B\/} {\bf
  64} 054403

\bibitem{Keren_SCGO}
Keren A, Uemura Y~J, Luke G, Mendels P, Mekata M and Asano T 2000 {\em Phys.
  Rev. Lett.\/} {\bf 84} 3450--3453

\bibitem{Bert_Tb2Sn2O7}
Bert F, Mendels P, Olariu A, Blanchard N, Collin G, Amato A, Baines C and
  Hillier A~D 2006 {\em Phys. Rev. Lett.\/} {\bf 97} 117203

\bibitem{Fukaya_volborthite}
Fukaya A, Fudamoto Y, Gat I~M, Ito T, Larkin M~I, Savici A~T, Uemura Y~J,
  Kyriakou P~P, Luke G~M, Rovers M~T, Kojima K~M, Keren A, Hanawa M and Hiroi Z
  2003 {\em Phys. Rev. Lett.\/} {\bf 91} 207603

\bibitem{Bert_volborthite}
Bert F, Mendels P, Bono D, Olariu A, Ladieu F, Trombe J~C, Duc F, Baines C,
  Amato A and Hillier A 2006 {\em Physica B\/} {\bf 374-375} 134--137

\bibitem{Lacroix_book}
Lacroix C, Mendels P and Mila F (eds) 2011 {\em Introduction to Frustrated
  Magnetism: Materials, Experiments, Theory\/} (Springer)

\end{thebibliography}

\end{document}